\documentclass{article}
\usepackage{amsfonts,amssymb, amsmath}
\usepackage[english]{babel}

\def\nn{\nonumber }
\def\bq{ \begin{equation}}
\def\eq{ \end{equation}}
\def\ben{ \begin{eqnarray}}
\def\en{ \end{eqnarray}}

\newtheorem{prop}{Proposition}

\newtheorem{re}{Remark}

\newtheorem{exa}{Example}

\begin{document}
%%%%%%%%%%%% TITLE %%%%%%%%%%%%%%

\title{New bi-Hamiltonian systems on the plane}
\author{A.V. Tsiganov \\
\it\small St.Petersburg State University, St.Petersburg, Russia\\
\it\small e--mail:  andrey.tsiganov@gmail.com}

\date{}
\maketitle

\begin{abstract}
 We discuss several  new bi-Hamiltonian  integrable systems  on the plane with integrals of motion of third, fourth and sixth order in momenta. The corresponding variables of separation, separated relations, compatible Poisson brackets and recursion operators are also presented  in the framework of the Jacobi method.
 \end{abstract}

\section{Introduction}
\setcounter{equation}{0}
The Jacobi method of separation of variables is a very important tool in analytical mechanics
Following  the classical work \cite{jac36},  the stationary Hamilton-Jacobi equation
\[H=E\]
is said to be separable in a set of canonical coordinates $u=(u_1,\ldots,u_m)$ and $p_u=(p_{u_1},\ldots,p_{u_m})$
\[
\{u_i,p_{u_j}\}=\delta_{ij}\,,\qquad  \{u_i,u_j\}=\{p_{u_i},p_{u_j}\}=0\,,\qquad i,j=1,\ldots,m\,,
\]
if there is an additively separated complete integral
 \[
 W(u_1,\ldots,u_m;\alpha_1,\ldots,\alpha_m)=\sum_{i=1}^m W_i(u_i;\alpha_1,\ldots,\alpha_m)\,,\qquad \alpha_1,\ldots,\alpha_m\in\mathbb R\,,
 \]
 depending non-trivially on a set of separation constants $\alpha_1,\ldots,\alpha_m$, and  $W_i$ are found in quadratures as solutions of ordinary differential equations.

In the framework of the Jacobi method, after finding  new variables of separation $u, p_u$ on  phase space $M$,  we have to look for  new problems to which they can be successfully applied, see lecture 26 in \cite{jac36}. Indeed, substituting canonical coordinates  into separated relations
\bq\label{gen-sep-rel}
\Phi_i(u_i,p_{u_i},H_1,\ldots, H_m)=0\,,\qquad i=1,\ldots,m, \qquad\mbox{with}\quad
\det\left[\dfrac{\partial \Phi_i}{\partial H_j}\right]\neq 0
\eq
and solving the resulting equations with respect to $H_1,\ldots,H_m$,  we obtain a new integrable system with independent  integrals of motion $H_1,\ldots,H_m$ in the involution.

In \cite{ts15a,ts15b,ts15c} we proposed to add one more step to this well-known  construction of integrable systems proposed by Jacobi.  Namely,  we can:
\begin{enumerate}
  \item take Hamilton-Jacobi equation $H=E$ separable in variables $u,p_u$;
  \item make auto B\"{a}cklund transformation (BT) of variables $(u,p_u)\to(\tilde{u},\tilde{p}_u)$, which conserves not only the Hamiltonian character of the equations of motion, but also the form of Hamilton-Jacobi equation;
  \item substitute new canonical variables $\tilde{u},\tilde{p}_{u}$ into the suitable separated relations  and  obtain new integrable systems.
\end{enumerate}
 The main aim of this paper is to extend this scheme by adding one more  step associated with a suitable change of time, which is similar to Weierstrass change of time for the Jacobi system on an ellipsoid   \cite{w}.

Below we take well-known Hamilton-Jacobi equations separable parabolic coordinates on  plane $Q=\mathbb R^2$
 \begin{equation}\label{u-par}
u_1 = q_2-\sqrt{q_1^2+q_2^2},\qquad u_2 = q_2+\sqrt{q_1^2+q_2^2}
\end{equation}
 and consider auto  B\"{a}cklund transformations $\mathcal B$ preserving the form of Hamilton and Hamilton-Jacobi equations.  After that, using  canonical variables on the phase space  $T^*\mathbb R^2$, we obtain  new integrable systems with the Hamilton function
 \bq\label{nat-ham}
H=\sum_{i,j=1}^2 \mathrm g_{ij}(u_1,u_2)p_{u_i}p_{u_j} +V(u_1,u_2)\,
\eq
  and second integrals of motion, which are polynomials of third, fourth and sixth order in momenta.  The corresponding vector fields are bi-Hamiltonian and, moreover, one of these systems is superintegrable.

 In classical mechanics auto BT of the Hamilton-Jacobi partial differential equation $H=E$  is canonical transformation of the phase space preserving the form of  this Hamilton-Jacobi equations  \cite{toda75,stef82}, i.e auto BTs are symmetries of the generic level set of integrals of motion. For many integrable by quadratures dynamical systems the generic level set is related to  the  Jacobian  of some hyperelliptic curve. It  allows us to study symmetries of integrals of motion using well-known  operations in  Jacobian
\bq\label{add-jac}
D\approx D',\quad  D+ D'=D''\qquad\mbox{and}\qquad [\ell] D=D''\,,
\eq
where $\approx$, $+$ and $[\ell]$ denotes equivalence, addition and scalar multiplication of divisors $D$ and $D'$, respectively.

Usually authors consider only addition of  full degree reduced divisor $D'$  and weight one reduced divisor  $D'$ in (\ref{add-jac}). This partial group operation and the corresponding  one-point auto B\"{a}cklund transformations have  been  studied from the different points of view in many publications, in particular see \cite{fed00,kuz02,skl00} and references therein. In fact, this  construction of auto BTs  was specially developed for  simplest one-parametric discretization of original continuous systems.

 Below we consider auto BTs associated with equivalence relation between two semi-reduced divisors $D$ and $D'$ which belong to a common equivalence class  in Jacobian using classical Abel results \cite{ab}, see also Abel's papers in  Crelle's Journal, v. \textbf{3} and v.\textbf{4}.  Such auto BTs represent hidden symmetries of the level manifold  which yield new canonical variables on original phase space and, therefore, we can use these new variables to  to construction of new integrable systems, i.e. to construction of hetero B\"{a}cklund transformations.

 The paper is organized as follows. In the remaining part of the first Section we show how
 our technique allows us to obtain well-known auto BTs which are related with translations  on the level manifold.  In Section 2 we apply the same  technique  to construct auto BT that represents a symmetry of the level manifold.  Of course, we can not apply  such auto BT to discretization of Hamiltonian system because when one iterates such BT one will obtain again original parabolic coordinates, but we can use such auto BT to construct new integrable systems. It is natural to use different types of auto BTs for the different purposes.

\subsection{Abel equations and B\"{a}cklund transformations}
According \cite{ab,bak97,gg, kl05}  Abel differential equation
\bq\label{ab-eq}
\omega\bigl(\mathrm p_1(t)\bigr)+\omega\bigl(\mathrm p_2(t)\bigr)+\cdots+\omega\bigl(\mathrm p_n(t)\bigr)=0
\eq
 on a hyperelliptic curve $\mathcal C$ of genus $g$
\[
\mathcal C:\qquad f(x,y)=y^2-f(x)=0\,,\qquad f(x)=a_{2g+2}x^{2g+2}+a_{2g+1}x^{2g+1}+\cdots a_0\,
\]
includes regular differential  $\omega$   at $n$ points $\mathrm p_k(t)$ on $\mathcal C$ moves with a rational parameter $t$. According to Abel's  idea, solutions  of  (\ref{ab-eq})  are  points $\mathrm p_k=(x_k,y_k)$ of   $\mathcal C$ intersecting with the second plane curve defined by equation
\[
h(x,y)=-y+P(x)=0\,.
\]
Here   $P(x)$ is the polynomial of $g+1$ order
 \[P(x)=b_{g+1}x^{g+1}+b_gx^g+\cdots+b_0\]
 with variable coefficients  $b_{g+1}\,,\ldots,b_0$ depending on $t$.  Eliminating $y$ from equations $f(x,y)=0$ and $h(x,y)=0$,  Abel determines abscissas $x_1,\ldots,x_n$ of the points  of intersection  as  roots of the Abel polynomial
\begin{equation}\label{abel-pol}
\psi(x)=f(x)-P^2(x)\,.
\end{equation}
If we suppose that $\psi(x)$ has no multiple  roots, then dividing roots of  the Abel polynomial $\psi$ (\ref{abel-pol}) in two parts  we can relate one subset of the roots to the other subset
\begin{equation}\label{ab-abs}
(x-x_1)\cdots(x-x_m)=\frac{f(x)-P(x)^2}{\phi(x)(x-x_{m+1})\cdots(x-x_n)}\,,
\end{equation}
 where $\phi(x)$ is a polynomial with independent on $t$ roots which play the role of constants of integration, see  discussion in textbook \cite{bak97}. According to  Abel \cite{ab}, polynomial $P(x)$ is usually constructed  using interpolation  because  ordinates of the points  of intersection are
\bq\label{ab-ord}
y_i=P(x_i)\,,\qquad i=1,\ldots,n.
\eq
Following to Euler,  mathematicians also investigated  algebraic integrals of the Abel equations (\ref{ab-eq}), which  are equivalent to the correspondence (\ref{ab-abs}-\ref{ab-ord})  \cite{bak97}.  For instance,  at $g=1$ there  are three possible Abel equations
\[
\dfrac{dx_1}{y_1}+\dfrac{dx_2}{y_2}=0\,,\qquad \dfrac{dx_1}{y_1}+\dfrac{dx_2}{y_2}+\dfrac{dx_3}{y_3}=0\,,\qquad
\dfrac{dx_1}{y_1}+\dfrac{dx_2}{y_2}+\dfrac{dx_3}{y_3}+\dfrac{dx_4}{y_4}=0
\]
on the symmetric products of two, three and four copies of   a compactified elliptic curve $\mathcal C$, respectively. These equations have  well-studied  algebraic integrals:
\[
\left(\frac{ y_1- y_2}{ x_1- x_2}\right)^2-a_4(  x_1+  x_2)^2-a_3( x_1+ x_2)=const\qquad \mbox{- the famous Euler integral}\,,\]
\[
 \frac{ y_i-y_k}{ x_i-x_k}-\sqrt{a_4} x_i= \frac{ y_j-y_k}{ x_j-x_k}-\sqrt{a_4} x_j\,,\qquad \mbox{- the  Abel integrals}
 \]
 and so-called complete integral
\bq\label{comp-int}
\left|
  \begin{array}{cccc}
     x_1^2 &  x_1 &1 &  y_1 \\
     x_2^2 &  x_2 &1 &  y_2 \\
     x_3^2 &  x_3 &1 &  y_3 \\
     x_4^2 &  x_4 & 1 & y_4 \\
  \end{array}
\right|=0\,,
\eq
respectively. These algebraic integrals relate two, three and  four points of intersection of the elliptic curve
$\mathcal C$ with  parabola \cite{bak97}.

Let us show how  well known  Abel relations (\ref{ab-abs}-\ref{ab-ord}) yield an auto B\"{a}cklund transformation of Hamilton-Jacobi equations. For instance, we take H\'{e}non-Heiles system with Hamiltonians
\begin{equation}\label{hh-1}
H_1=\frac{p_1^2+p_2^2}{4}-4aq_2(q_1^2+2q_2^2)\,,\qquad
H_2=\frac{p_1(q_1p_2-q_2p_1)}{2}-aq_1^2(q_1^2+4q_2^2)\,
\end{equation}
separable in parabolic coordinates on the plane (\ref{u-par}). To describe evolution of $u_ {1,2} $ with respect to $H_ {1,2} $ we use the canonical Poisson bracket
\bq\label{poi1}
\{u_i,p_{u_j}\}=\delta_{ij}\,,\quad \{u_1,u_2\}=\{p_{u_1},p_{u_2}\}=0\,.
\eq
and expressions for $H_{1,2}$
\bq\label{hh-ham-u}
H_1=\frac{\scriptstyle p_{u_1}^2u_1-p_{u_2}^2u_2}{\scriptstyle u_1-u_2}-a(u_1+u_2)(u_1^2+u_2^2)\,,\quad
H_2=\frac{\scriptstyle u_1u_2(p_{u_1}^2-p_{u_2}^2)}{\scriptstyle u_2-u_1}+au_1u_2(u_1^2+u_1u_2+u_2^2)
\eq
 to obtain
\begin{equation}\label{hh-eq1}
\frac{du_1}{dt_1}=\{u_1,H_1\}=\frac{2p_{u_1}u_1}{u_1-u_2}\,,\qquad
\frac{du_2}{dt_1}=\{u_2,H_1\}=\frac{2p_{u_2}u_2}{u_2-u_1}
\end{equation}
and
\begin{equation}\label{hh-eq2}
\frac{du_1}{dt_2}=\{u_1,H_2\}=\frac{2u_1u_2p_{u_1}}{u_2-u_1}\,,\qquad
\frac{du_2}{dt_2}=\{u_2,H_2\}=\frac{2u_1u_2p_{u_2}}{u_1-u_2}\,.
\end{equation}
Here $p_{u_{1,2}}$ are the standard momenta associated with parabolic coordinates $u_{1,2}$:
\[
p_{u_1}=\frac{p_2}{2}-\frac{p_1(q_2+\sqrt{q_1^2+q_2^2})}{2q_1}\,,\qquad
p_{u_2}=\frac{p_2}{2}-\frac{p_1(q_2-\sqrt{q_1^2+q_2^2})}{2q_1}\,.
\]
Using Hamilton-Jacobi equations $H_{1,2}=\alpha_{1,2}$ we can prove that these variables satisfy to the following separated relations
\begin{equation}\label{hh-sep}
\bigl(u_ip_{u_i}\bigr)^2=u_i(au_i^4+\alpha_1u_i+\alpha_2)\,,\qquad i=1,2.
\end{equation}
Expressions (\ref{hh-eq1}-\ref{hh-eq2} ) and (\ref{hh-sep}) yield standard Abel quadratures
\bq\label{hh-ab-q1}
\frac{du_1}{\sqrt{f(u_1)}}+\frac{du_2}{\sqrt{f(u_2)}}=2dt_2\,,\qquad \frac{u_1du_1}{\sqrt{f(u_1)}}+\frac{u_2du_2}{\sqrt{f(u_2)}}=2dt_1,
\eq
on the  hyperelliptic curve $\mathcal C$ of genus two  defined by equation
\bq\label{hh-eq-c}
f(x,y)=y^2-f(x)=0\,,\qquad f(x)=x(ax^4+\alpha_1x+\alpha_2)\,.
\eq

Suppose that transformation of variables
\bq\label{st-b-hh}\mathcal B:\quad (u_{1},u_2,p_{u_1},p_{u_2})\to
(\tilde{u}_{1},\tilde{u}_2,\tilde{p}_{u_1},\tilde{p}_{u_2})
\eq
preserves Hamilton equations (\ref{hh-eq1}-\ref{hh-eq2}) and the form of Hamiltonians (\ref{hh-ham-u}). It means that new variables satisfy to the same equations
\bq\label{hh-ab-q2}
\frac{d\tilde{u}_1}{\sqrt{f(\tilde{u}_1)}}+\frac{d\tilde{u}_2}{\sqrt{f(\tilde{u}_2)}}=2dt_2\,,\qquad \frac{\tilde{u}_1d\tilde{u}_1}{\sqrt{f(\tilde{u}_1)}}+\frac{\tilde{u}_2d\tilde{u}_2}{\sqrt{f(\tilde{u}_2)}}=2dt_1\,.
\eq
Subtracting (\ref{hh-ab-q2}) from (\ref{hh-ab-q1}) one gets two Abel differential equations (\ref{ab-eq})
\begin{equation}\label{ab-eq-g2}
\begin{array}{c}
\omega_1(x_1,y_1)+\omega_1(x_2,y_2)+\omega_1(x_3,y_3)+\omega_1(x_4,y_4)=0\,,\\ \\
\omega_2(x_1,y_1)+\omega_2(x_2,y_2)+\omega_2(x_3,y_3)+\omega_2(x_4,y_4)=0\,,
\end{array}
\end{equation}
where
 \[
 x_{1,2}=u_{1,2},\quad y_{1,2}=u_{1,2}p_{u_{1,2}}\,,\qquad
x_{3,4}=\tilde{u}_{1,2},\quad y_{3,4}=-\tilde{u}_{1,2}\tilde{p}_{u_{1,2}}
\]
and $\omega_{1,2}$ form a base of holomorphic differentials on hyperelliptic curve $\mathcal C$ of genus $g=2$
\[
\omega_1(x,y)=\frac{dx}{y}\,,\qquad \omega_2(x,y)=\frac{xdx}{y}\,.
\]
These Abel equations are closely related to the so-called Picard group which is isomorphic to Jacobian of $\mathcal C$ \cite{kl05}.

According \cite{bak97} there are six points of intersection:
\begin{enumerate}
   \item one point can be taken at infinity;
  \item four points  $(x_1,y_1),\ldots,(x_4,y_4)$ are solutions of Abel equations (\ref{ab-eq-g2}), which form support of weigh two divisors $D$ and $D''$, respectively;
   \item one remaining point with coordinates $ (\lambda,\mu)$ is independent on $t$ and belong to support of weight one divisor $D'$.
\end{enumerate}
Abscissa $\lambda$ is an arbitrary constant of integration of Abel differential equations (\ref{ab-eq-g2}).

Using the corresponding Abel polynomial
 \[
\psi(x)=f(x)-P(x)^2=a(x-\lambda)(x-x_1)(x-x_2)(x-x_3)(x-x_4)
\]
we can determine new variables $x_{3,4}=\tilde{u}_{1,2}$ as functions on initial variables $x_{1,2}$ and $y_{1,2}$
\bq\label{new-var-hh}
(x-x_3)(x-x_4)=\dfrac{f(x)-P(x)^2}{a(x-x_1)(x-x_2)(x-\lambda)}.
\eq
In our case the number of degrees of freedom  $m=2$ is equal to the genus $g=2$  and, therefore, we have
\[
P(x)=
\dfrac{y_1(x-x_2)(x-\lambda)}{(x_1-x_2)(x_1-\lambda)}+\dfrac{y_2(x-x_1)(x-\lambda)}{(x_2-x_1)(x_2-\lambda)}
+\dfrac{\mu(x-x_1)(x-x_2)}{(\lambda-x_1)(\lambda-x_2)}
\]
due to the Lagrange interpolation formulae.

Using (\ref{new-var-hh}) together with definition of ordinates $y_{3,4}=P(x_{3,4})$ (\ref{ab-ord}) we can explicitly obtain the desired transformation $\mathcal B$ (\ref{st-b-hh}) which can be easy  rewritten as  canonical transformation of original coordinates on $T^*\mathbb R^2$
\[\begin{array}{l}
\tilde{q}_1^2=q_1^2+\frac{8aq_1^2(2\lambda q_2+q_1^2+2q_2^2)+p_1(\lambda p_1-2p_1q_2+2p_2q_1)}{4a(\lambda^2-2\lambda q_2-q_1^2)}+\frac{q_1\bigl(\lambda q_1p_2^2+2q_1p_2(p_1q_1-\mu)+p_1(\lambda-2q_2)(p_1q_1-2\mu)\bigr)}
{2a(\lambda^2-2\lambda q_2-q_1^2)^2}
\\ \\
\tilde{q}_2=q_2+\frac{8a(q_1^2+6q_2^2)\lambda+16aq_1^2q_2+p_1^2+2p_2^2}{8a(\lambda^2-2\lambda q_2-q_1^2)}-
\frac{2p_2(\mu-p_1q_1-p_2q_2)\lambda+q_1\bigl(2p_1\mu -q_1(p_1^2+p_2^2)\bigr)}
{4a(\lambda^2-2\lambda q_2-q_1^2)^2}
\end{array}
\]
and
\[
\tilde{q_1}\tilde{p}_1=-p_1q_1+\dfrac{(q_1^2-\tilde{q}_1^2)(2\mu-\lambda p_2-p_1q_1)}{\lambda^2-2\lambda q_2-q_1^2}\,,
\quad
\tilde{p}_2=-p_2+\frac{2(q_2-\tilde{q}_2)(2\mu-\lambda p_2-p_1q_1)}{\lambda^2-2\lambda q_2-q_1^2}\,.
\]
Auto B\"{a}cklund transformation  $\mathcal B$ (\ref{st-b-hh}) is the hidden symmetry of the level set of integrals of motion for any $\lambda$. It allows us to apply this canonical transformation $\mathcal B$ (\ref{st-b-hh}) to  construction of new integrable systems  in the framework of the Jacobi method. Indeed, if  $\lambda=0$ in (\ref{new-var-hh}), variables $\tilde{u}_{1,2}$ are the roots of   polynomial
\[
(x-\tilde{u}_1)(x-\tilde{u}_2)=x^2+\left(u_1+u_2-\dfrac{(p_{u_1}-p_{u_2})^2}{a(u_1-u_2)^2}\right)x+u_1^2+u_1u_2+u_2^2-\dfrac{p_{u_1}^2-p_{u_2}^2}{a(u_1-u_2)}\,.
\]
whereas momenta are equal to
\[
\tilde{p}_{1,2}={\tilde{u}_{1,2}^{-1}}\,P(\tilde{u}_{1,2})\,,\qquad P(x)=\dfrac{x(x-u_2)}{u_1-u_2}p_{u_1}+\dfrac{x(x-u_1)}{u_2-u_1}p_{u_2}\,.
\]
 It is easy to prove that original Poisson bracket (\ref{poi1}) in these variables reads as
\[
\{\tilde{u}_i,\tilde{p}_{u_j}\}=\delta_{ij}\,,\quad \{\tilde{u}_1,\tilde{u}_2\}=\{\tilde{p}_{u_1},\tilde{p}_{u_2}\}=0\,.
\]
Substituting these new canonical variables on $T^*\mathbb R^2$ into a pair of separation relations
\bq\label{hh-sep-rel2}
\Bigl(\tilde{p}^2_{u_{i}}-a\tilde{u}_{i}^3\Bigr)=\tilde{H}_1\pm\sqrt{\tilde{H}_2}\,,\qquad i=1,2
\eq
and solving the resulting equations with respect to $\tilde{H}_{1,2}$, one gets the well-known Hamilton function
\bq\label{second-hh}
\tilde{H}_1=\Bigl(\tilde{p}^2_{u_{1}}-a\tilde{u}_{1}^3\Bigr)+\Bigl(\tilde{p}^2_{u_{2}}-a\tilde{u}_{2}^3\Bigr)=\dfrac{p_1^2}{4}+\dfrac{p_2^2}{2}-2aq_2(3q_1^2+8q_2^2)
\eq
for another integrable case of the  H\'{e}non-Heiles system. Second integral of motion $\tilde{H}_2$ is a polynomial of fourth order in momenta $p_{1,2}$, see details in \cite{ts15c}.

 In this case the number of degrees of freedom  $m=2$ is equal to the genus $g=2$ of hyperelliptic curve $\mathcal C $
 and relations (\ref{ab-abs}-\ref{ab-ord})  determine group operations on the Jacobian of $\mathcal C$. Since the Jacobian of a hyperelliptic curve is the group of degree zero divisors modulo principal divisors, the group operation is formal addition modulo the equivalence relation. Sequently we can say that auto B\"{a}cklund transformation (\ref{st-b-hh})  is a shift on the Jacobi variety of $\mathcal C$ which can be described using  the Mumford coordinates of divisors \cite{kuz02,mum}.

 Indeed,   we can introduce the Jacobi polynomials \cite{mum}:
\bq\label{uvw-pol}
U(x)=\prod_{k=1}^m (x-  x_k)\,,\quad
V(x)=\displaystyle \sum_{i=1}^m   y_i\prod_{j\neq i}\dfrac{x-  x_j}{  x_i-  x_j}\,,\quad
W(x)=\dfrac{f(x)-V^2(x)}{U(x)}
\eq
and  $2\times 2$ Lax matrix for the  H\'{e}non-Heiles system
\bq\label{jac-lax}
L(x)=\left(
       \begin{array}{cc}
         V(x) & U(x) \\
         W(x) & -V(x)\\
       \end{array}
     \right)\,.
\eq
 After  similar  transformation
\[
L(x)\to \tilde{L}(x)=ML(x)M^{-1}\,,\qquad
M=\left(
    \begin{array}{cc}
      U(x) & 0 \\
    V(x)-P(x) & U(x) \\
    \end{array}
  \right)
\]
one gets Lax matrix
\bq\label{ab-lax}
\tilde{L}(x)=\left(
               \begin{array}{cc}
                 P(x) & U(x) \\
                \psi(x)U^{-1} &- P(x) \\
               \end{array}
             \right)
\eq
defined by Abel polynomials $\psi(x)$, $P(x)$  and $U(x)$.  At $\lambda=0$ this Lax matrix was obtained in \cite{ts15a,ts15c}.

\section{Integrable system with integral of motion of sixth order in momenta}
\setcounter{equation}{0}
Let us take an integrable system with Hamiltonians
\bq\label{ham-par}
\begin{array}{l}
H_1=\dfrac{p_1^2+p_2^2}{4}+\dfrac{a}{q_1^2}-\dfrac{2bq_2}{q_1^4}+\dfrac{c(q_1^2+4q_2^2)}{q_1^6}\,,\qquad a,b,c\in\mathbb R\,,
\\ \\
H_2=-\dfrac{p_1(p_1q_2-p_2q_1)}{2}-\dfrac{2aq_2}{q_1^2}+\dfrac{b(q_1^2+4q_2^2)}{q_1^4}-\dfrac{4cq_2(q_1^2+2q_2^2)}{q_1^6}\,.
\end{array}
\eq
According to \cite{ts15}  there is integrable deformation  of this Hamilton function at $b=0$
\[
\bar{H}_1=H+\Delta H=\dfrac{p_1^2+p_2^2}{4}+\dfrac{a}{q_1^2}+\dfrac{c(q_1^2+4q_2^2)}{q_1^6}+d(q_1^2+q_2^2)+\dfrac{e}{q_2^2}\,.\nn\\
\]
The corresponding second integral of motion is the polynomial of fourth order in momenta.

Our initial aim was to find variables of separation for this system in the framework of  Abel theory. Instead of this, we find new integrable deformation of the Hamiltonian (\ref{ham-par}) with second polynomial integrals of sixth order in momenta. Below we describe construction of this new integrable system in details.

\subsection{Separation of variables and Abel differential equation}
In parabolic coordinates Hamiltonians $H_{1,2}$ (\ref{ham-par})  look like
 \bq
 \label{ham-uu}
 \begin{array}{l}
 H_1=\dfrac{u_1p_{u_1}^2}{u_1-u_2}+\dfrac{u_2p_{u_2}^2}{u_2-u_1} -\dfrac{a}{u_1u_2}-\dfrac{b(u_1+u_2)}{u_1^2u_2^2}-\dfrac{c(u_1^2+u_1u_2+u_2^2)}{u_1^3u_2^3}\,,
 \\
\\
H_2=\dfrac{u_1 u_2 p_{u_1}^2}{u_2-u_1}+\dfrac{ u_1 u_2p_{u_2}^2}{u_1-u_2}+\dfrac{a(u_1+u_2)}{u_1 u_2}
+\dfrac{b(u_1^2+u_1 u_2+u_2^2) }{u_1^2 u_2^2}+\dfrac{c(u_1+u_2) (u_1^2+u_2^2) }{u_1^3 u_2^3}\,.
\end{array}
\eq
The corresponding Hamilton-Jacobi equations $H_{1,2}=\alpha_{1,2}$ are equivalent to  separation relations
\[
\Phi(u_i,p_i,H_1,H_2)=\left(u_i^2p_{u_i}\right)^2-(H_1u_i^4+H_2u_i^3-au_i^2-bu_i-c)=0 \,,
\]
which determine the elliptic curve
\bq\label{ell-par}
\mathcal C:\qquad y^2=f(x)\,,\qquad f(x)=\alpha_1u^4+\alpha_2u^3-au^2-bu-c\,,
\eq
i.e. variables
\[x_{1,2}=u_{1,2},\qquad y_{1,2}=u_{1,2}^2p_{u_{1,2}}\]
may be identified with abscissas and ordinates of two points on  $\mathcal C$. In this case the number degrees of freedom $m=2$ does no equal to the genus $g=1$ of the underlying elliptic curve and, therefore,  standard construction of BTs from \cite{fed00,kuz02}
cannot be applied because   it works only when $m = g$.

Equations of motion for  $x_{1,2}$ are equal to
\bq\label{flow1}
\dfrac{dx_1}{dt}=\{u_1,H_1\}=\dfrac{\partial H_1}{\partial p_{u_1}}=\dfrac{2y_1}{x_1(x_1-x_2)}\,,
\qquad
\dfrac{dx_2}{dt}=\{u_2,H_1\}=\dfrac{\partial H_1}{\partial p_{u_2}}=\dfrac{2y_2}{x_2(x_2-x_1)}\,.
\eq
It allows us to obtain Abel quadratures
\[
 \dfrac{x_1dx_1}{y_1}+ \dfrac{x_2dx_2}{y_2}=0\,,\qquad\mbox{and}\qquad
 \dfrac{x_1^2dx_1}{y_1}+\dfrac{x_2^2 dx_2}{y_2}=2dt\,,
\]
which we have to solve with respect to functions $x_{1,2}(t)$ on time \cite{jac36}.  In order to get these functions
 we can change time  $t \to s$ in (\ref{flow1})  following  to Weierstrass idea  \cite{w} and introduce new equations
 \bq\label{flow11}
\dfrac{du_1}{ds}=\{u_1,H_1\}_W=\dfrac{2x_1y_1}{x_1-x_2}\,,
\qquad
\dfrac{du_2}{ds}=\{u_2,H_1\}_W=\dfrac{2x_2y_2}{x_2-x_1}\,.
\eq
 in order to reduce Abel quadratures to the following  form
 \bq\label{ab-qv2}
 \dfrac{dx_1}{x_1y_1}+ \dfrac{dx_2}{x_2y_2}=0\,,\qquad
 \dfrac{dx_1}{y_1}+\dfrac{dx_2}{y_2}=2ds_1\,.
\eq
After the Weierstrass change of time second quadrature incorporates standard holomorphic differential on the elliptic curve $\mathcal C$ that allows us to relate this equation with Jacobian \cite{kl05}.  Equations of motion (\ref{flow11}) are  Hamiltonian equations with respect to the new Poisson bracket
\bq\label{poi2}
\{u_i,p_{u_j}\}_W=u_i^2\delta_{ij}\,,\qquad  \{u_1,u_2\}_W=\{p_{u_1},p_{u_2}\}_W=0\,,
\eq
which is compatible with the original canonical bracket $\{.,.\}$  (\ref{poi1}).

Suppose that transformation of variables
 \bq\label{st-b}
\mathcal  B:\qquad (u_1,u_2,p_{u_1},p_{u_2})\leftrightarrow (\tilde{u}_1,\tilde{u}_2,\tilde{p}_{u_1},\tilde{p}_{u_2})
\eq
preserves form of Hamilton equations  (\ref{flow11}) and Hamilton-Jacobi equations, i.e.
\[
H_1(u,p_u)=\alpha_1=H_1(\tilde{u},\tilde{p}_{u_1})\,,\qquad
H_2(u,p_u)=\alpha_2=H_2(\tilde{u},\tilde{p}_{u_1})\,,
\]
 where functions $H_{1,2}$ are given by (\ref{ham-uu}).  It means that variables  $x_{1,2}, y_{1,2}$ and
 \[
x_{3,4}=\tilde{u}_{1,2}\,,\qquad y_{3,4}=-\tilde{u}_{1,2}^2\tilde{p}_{u_{1,2}}\,.
\]
 satisfy to  the  Abel differential  equation
\bq\label{eq-ab}
 \dfrac{dx_1}{y_1}+\dfrac{dx_2}{y_2}+\dfrac{dx_3}{y_3}+\dfrac{dx_4}{y_4}=0\,.
\eq
In this case,  points on the curve with coordinates $(x_{1,2}y_{1,2})$ and $(x_{3,4},y_{3,4})$ form weight two unreduced divisors $D$ and $D'$ on $\mathcal C$. An equivalence relations between these divisors we identify with desired auto BT. Indeed,  the corresponding Abel polynomial is equal to
  \[
\psi(x)=f(x)-P(x)^2=A(x-x_1)(x-x_2)(x-x_3)(x-x_4)\,,\qquad A=\alpha_1-b_2^2\,,
\]
where $b_2$ is  the coefficient of polynomial $P(x)=b_2x^2+b_1x+b_0$. It allows us to explicitly determine the desired
mapping  $\mathcal B$ (\ref{st-b}) using polynomial
\begin{equation}\label{xy-h}
(x-x_3)(x-x_4)=\frac{f(x)-P(x)^2}{A(x-{x}_1)(x-x_2)}\,,
\end{equation}
where $f(x)$ is given by (\ref{ell-par}),   and polynomial of the second order
 \begin{equation}\label{p-ab-lag}
P(x)=x\left(\dfrac{(x-x_2)y_1}{x_1(x_1-x_2)}+\dfrac{(x-x_1)y_2}{x_2(x_2-x_1)}\right)
\end{equation}
 which is completely defined by its values at $x_k$
 \[
 y_k=P(x_k)\,,\qquad k=1,\ldots,4,
 \]
 and by  well known algebraic integral (\ref{comp-int}).

\begin{prop}
If  variables $\tilde{u}_{1,2}$ are solutions of equation
\[
\left(H_1-\left(\dfrac{u_1p_{u_1}-u_2p_{u_2}}{u_1-u_2}\right)^2\right)x^2-\dfrac{bu_1u_2+c(u_1+u_2)}{u_1^2u_2^2}\,x-\dfrac{c}{u_1u_2}=0
\]
and
\[
\left.\tilde{p}_{u_{1,2}}=-\tilde{u}_{1,2}^{-2}\,P\right|_{x=\tilde{u}_{1,2}}\,,\quad P(x)=x\left(\dfrac{(x-u_2)u_1p_{u_1}}{u_1-u_2}+\dfrac{(x-u_1)u_2p_{u_2}}{u_2-u_1}\right)\,,
\]
then mapping $\mathcal B$ (\ref{st-b}) preserves  the form of Hamilton equations  (\ref{flow11}),  the form of Hamiltonians  $H_{1,2}$ (\ref{ham-par}-\ref{ham-uu}) and the form of the Poisson bracket $\{.,.\}_W$ (\ref{poi2}), i.e.
\[
\{\tilde{u}_i,\tilde{p}_{u_j}\}_W=\tilde{u}_i^2\delta_{ij}\,,\qquad  \{\tilde{u}_1,\tilde{u}_2\}_W=\{\tilde{p}_{u_1},\tilde{p}_{u_2}\}_W=0\,.
\]
\end{prop}
The proof is a straightforward calculation.

Of course, we can not use this mapping $\mathcal B$ (\ref{st-b}) for discretization of Hamiltonian flows (\ref{flow11}), but this auto BT is  the hidden symmetry which describes fundamental properties of the given Hamiltonian system similar to the Noether symmetries.

  \subsection{Construction of the new integrable system on $T^*\mathbb R$}
The following Poisson map
 \[
 \rho:\qquad (u_1,u_2,p_{u_1},p_{u_2})\to (u_1,u_2,u_1^2p_{u_1},u_2^2p_{u_2})
 \]
  reduces canonical Poisson bracket $\{.,.\}$ to  bracket $\{.,.\}_W$, which allows us to rewrite  equations of motion (\ref{flow11}) in Hamiltonian form.

  Using the composition of  Poisson mappings  $\rho$ and  $\mathcal B$  (\ref{st-b}) we determine variables
 $\hat{u}_{1,2}$, which are solutions of equation
  \bq\label{new-u2}
  \left(\rho(H_1)-\left(\dfrac{u_1^3p_{u_1}-u_2^3p_{u_2}}{u_1-u_2}\right)^2\right)x^2-\dfrac{bu_1u_2+c(u_1+u_2)}{u_1^2u_2^2}\,x-\dfrac{c}{u_1u_2}=0
  \eq
  where
 \[
  \rho(H_1)=\dfrac{u_1^5p_{u_1}^2}{u_1-u_2}+\dfrac{u_2^5p_{u_2}^2}{u_2-u_1} -\dfrac{a}{u_1u_2}-\dfrac{b(u_1+u_2)}{u_1^2u_2^2}-\dfrac{c(u_1^2+u_1u_2+u_2^2)}{u_1^3u_2^3}\,.
 \]
The corresponding momenta are equal to
\bq\label{new-p2}
\left.\hat{p}_{u_{1,2}}=-\hat{u}_{1,2}^{-4}\,\rho(P)\right|_{x=\hat{u}_{1,2}}\,,\quad \rho(P)=x\left(\dfrac{(x-u_2)u_1^3p_{u_1}}{u_1-u_2}+\dfrac{(x-u_1)u_2^3p_{u_2}}{u_2-u_1}\right)\,.
\eq
Straightforward calculation allows us to prove the following statement.
\begin{prop}
Canonical Poisson bracket
\[
\{u_i,p_{u_j}\}=\delta_{ij}\,,\qquad  \{u_1,u_2\}=\{p_{u_1},p_{u_2}\}=0
\]
 has the same form
\[
\{\hat{u}_i,\hat{p}_{u_j}\}=\delta_{ij}\,,\qquad  \{\hat{u}_1,\hat{u}_2\}=\{\hat{p}_{u_1},\hat{p}_{u_2}\}=0
\]
in variables  $\hat{u}_{1,2}$ and  $\hat{p}_{u_{1,2}}$  (\ref{new-u2},\ref{new-p2}).
\end{prop}
Thus, we obtain new canonical variables on  phase space $T^*\mathbb R^2$, which can be useful to construction of new integrable systems in the frameworks of  the Jacobi method.

For instance, let us substitute these variables into the separated relations similar to (\ref{hh-sep-rel2})
\bq
\label{n1-sep-rel}
2\lambda_i
=\left(\hat{u}_i^4\hat{p}_{u_i}^2+\dfrac{a}{\hat{u}_i^2}+\dfrac{b}{\hat{u}_i^3}+\dfrac{c}{\hat{u}_i^4}\right)=\hat{H}_1\pm\sqrt{\hat{H}_2}\,,\qquad i=1,2,
\eq
and solve these relations with respect to $\hat{H}_{1,2}$.
\begin{prop}
Functions on  phase space $T^*\mathbb R^2$
\[
\hat{H}_1=\lambda_1+\lambda_2\,,\qquad \hat{H}_2=(\lambda_1-\lambda_2)^2\,,
\]
are in involution with respect to the following compatible Poisson brackets
\bq\label{poi-11}
\{\hat{u}_i,\hat{p}_{u_j}\}=\delta_{i,j}\,,\qquad \{\hat{u}_1,\hat{u}_2\}=\{\hat{p}_{u_1},\hat{p}_{u_2}\}=0\,,
\eq
 and
\bq\label{poi-22}
\{\hat{u}_i,\hat{p}_{u_j}\}'=\lambda_i^{-1}\delta_{i,j}\,,\qquad \{\hat{u}_1,\hat{u}_2\}'=\{\hat{p}_{u_1},\hat{p}_{u_2}\}'=0\,.
\eq
\end{prop}
The proof is a  straightforward calculation.

Moreover, using the corresponding bivectors $P$ and $P'$ it is easy to prove that vector field
\[X=Pd(\lambda_1+\lambda_2)=P'd\left(\dfrac{\lambda_1^2+\lambda_2^2}{2}\right)
\]
is bi-Hamiltonian vector field.  This trivial in $\hat{u}_{1,2}$ and $\hat{p}_{u_{1,2}}$ variables Hamiltonian $\hat{H}_1=\lambda_1+\lambda_2$ has more complicated form in original parabolic coordinates and momenta:
\[\begin{array}{l}
\hat{H}_1=\dfrac{\bigl(bu_1u_2+c(3u_1+u_2)\bigr)u_1^4p_{u_1}^2}{c(u_1-u_2)}
+\dfrac{\bigl(bu_1u_2+c(u_1+3u_2)\bigr)u_2^4p_{u_2}^2}{c(u_2-u_1)}-\dfrac{(bu_1+c)(au_1^2+bu_1+c)}{cu_1^4}\\
\\
-\dfrac{(bu_2+c)(au_2^2+bu_2+c)}{cu_2^4}
-\dfrac{4ac+b^2}{cu_1u_2}-\dfrac{5b(u_1+u_2)}{u_1^2u_2^2}-\dfrac{4c(u_1^2+u_1u_2+u_2^2)}{u_1^3u_2^3}\,.
\end{array}
\]
Second integral is  polynomial of  sixth order in momenta
\[\begin{array}{l}
\hat{H}_2=\left(cu_1^4u_2^4(u_1^2p_{u_1}-u_2^2p_{u_2})^2+\frac{(u_1-u_2)^2\Bigl((3u_1^2+2u_1u_2+3u_2^2)c^2+2u_1u_2\bigl(2au_1u_2+b(u_1+u_2)\bigr)c-b^2u_1^2u_2^2\Bigr)}{4}
\right)\\
\\
\phantom{\hat{H}_2}\times\left(
\dfrac{u_1^4p_{u_1}^2-u_2^4p_{u_2}^2}{c(u_1-u_2)^2}-\dfrac{a(u_1+u_2)}{cu_1^2u_2^2(u_1-u_2)}
-\dfrac{b(u_1^2+u_1u_2+u_2^2)}{cu_1^3u_2^3(u_1-u_2)}
-\dfrac{(u_1+u_2)(u_1^2+u_2^2)}{u_1^4u_2^4(u_1-u_2)}
\right)^2
\end{array}
\]
 If we put   $a=b=0$ and then $c=0$, the second integral of motion $cH_2=K^2$ becomes a complete square. So, we have the geodesic flow on the plane with integrals of motion
\[
\hat{H}_1=T=\dfrac{ u_1^4(3u_1+u_2)p_{u_1}^2}{u_1-u_2}+\dfrac{u_2^4(u_1+3u_2)p_{u_2}^2}{u_2-u_1} \,,
\qquad
K=\dfrac{u_1^2u_2^2(u_1^2p_{u_1}-u_2^2p_{u_2})(u_1^4p_{u_1}^2-u_2^4p_{u_2}^2)}{(u_1-u_2)^2} \,.
\]
In  Cartesian coordinates   these integrals of motion read as
\[
T=\dfrac{(q_1^2+6q_2^2)q_1^2p_1^2}{2}+(5q_1^2+12q_2^2)q_1q_2p_1p_2+\dfrac{(q_1^4+8q_1^2q_2^2+12q_2^4)p_2^2}{2}
\]
and
\[
K=\dfrac{q_1^7q_2p_1^3}4+\dfrac{q_1^6(q_1^2+6q_2^2)p_1^2p_2}{4}+q_1^5q_2(q_1^2+3q_2^2)p_1p_2^2+q_1^4(q_1^2+2q_2^2)q_2^2p_2^3\,.
\]
It is easy to prove that Hamiltonian $T$ has no polynomial integrals of motion of  first or second order in momenta.

Such nonstandard Hamiltonians  may appear in the study of a wide range of fields such  as nonholonomic dynamics, control theory,  seismology, biology, in the study of a self graviting stellar gas cloud, optoelectronics, fluid mechanics etc.

\section{Other new integrable systems on the plane}
\setcounter{equation}{0}
If we substitute parabolic coordinates $u_{1,2}$ and conjugated momenta $p_{u_{1,2}}$ into a family of separated relations

\bq\label{eq-metr}
\begin{array}{llcl}
A:\qquad& \left(u^2p_{u}\right)^2-(H_1u^4+H_2u^3+au^2+bu+c)&=&0 \,,\\
B:\qquad &\left(u^2p_{u}\right)^2-(au^4+H_1u^3+H_2u^2+bu+c)&=&0\,,\\
C:\qquad &\left(u^2p_{u}\right)^2-(au^4+bu^3+H_1u^2+H_2u+c)&=&0\,,\\
D:\qquad &\left(u^2p_{u}\right)^2-(au^4+bu^3+cu^2+H_1u+H_2)&=&0
\end{array}
\eq
and solve the resulting pairs of equations with respect to $H_{1,2}$, we obtain dual St\"{a}ckel systems for which every trajectory of one system is a reparametrized trajectory of the other system\cite{ts00}. The corresponding integrable diagonal metrics
\[
\mathrm g_{km}=\left(
              \begin{array}{cc}
                \dfrac{u_2^ku_1^m}{u_1-u_2}& 0 \\
                0 & \dfrac{u_1^ku_2^m}{u_2-u_1} \\
              \end{array}
            \right)\,,\qquad   k=0,1;\quad m=1,\ldots,4,
\]
are geodesically  equivalent metrics \cite{mt01}.

For each of these systems we can construct  an analogue of the B\"{a}cklund transformation and  Poisson map $\rho$, associated with the Weierstrass change of time, that allows us  to get different canonical variables and different integrable systems on $T^*\mathbb R^2$.
The first case in (\ref{eq-metr}) was considered in the previous Section, whereas the third case  leads to an integrable system with quadratic integrals of motion. Thus, below we consider only the second and fourth separation relations in (\ref{eq-metr}).

\subsection{Case B.}
In this case
\[
f(x)= ax^4+H_1x^3+H_2x^2+bx+c\,,\quad
P(x)=x\left(\dfrac{(x-x_2)y_1}{x_1(x_1-x_2)}+\dfrac{(x-x_1)y_2}{x_2(x_2-x_1)}\right)
\]
and  variables $x_{3,4}=\tilde{u}_{1,2}$ are the roots of polynomial
\[
\dfrac{\psi(x)}{(x-u_1)(x-x_2)}=
\left(a-\dfrac{(u_1p_{u_1}-u_2p_{u_2})^2}{(u_1-u_2)^2}\right)x^2+\dfrac{bu_1u_2+c(u_1+u_2)}{u_1^2u_2^2}\,x
+\dfrac{c}{u_1 u_2}\,.
\]
These coordinates commute with respect to the Poisson brackets
\[
\{u_i,p_{u_j}\}_W=u_i\delta_{ij}\,,\qquad  \{u_1,u_2\}_W=\{p_{u_1},p_{u_2}\}_W=0\,.
\]
 Using an additional Poisson map
\[
\rho_B:\qquad (u_1,u_2,p_{u_1},p_{u_2})\to (u_1,u_2,u_1p_{u_1},u_2p_{u_2})\,,\]
we can define canonical  variables $\hat{u}_{1,2}$ on $T^*\mathbb R^2$,  which are  the roots of polynomial
\[
\rho_B\left(\dfrac{\psi(x)}{(x-u_1)(x-x_2)}\right)=
\left(a-\dfrac{(u_1^2p_{u_1}-u_2^2p_{u_2})^2}{(u_1-u_2)^2}\right)x^2+\dfrac{bu_1u_2+c(u_1+u_2)}{u_1^2u_2^2}\,x
+\dfrac{c}{u_1 u_2}\,,
\]
and the conjugated momenta
\[
\left.\hat{p}_{u_{1,2}}=-\hat{u}_{1,2}^{-3}\,\rho_B(P)\right|_{x=\hat{u}_{1,2}}\,,\quad \rho_B(P)=x\left(\dfrac{(x-u_2)u_1^2p_{u_1}}{u_1-u_2}+\dfrac{(x-u_1)u_2^2p_{u_2}}{u_2-u_1}\right)\,,
\]
so that
\[
\{\hat{u}_i,\hat{p}_{u_j}\}=\delta_{ij}\,,\qquad  \{\hat{u}_1,\hat{u}_2\}=\{\hat{p}_{u_1},\hat{p}_{u_2}\}=0\,.
\]
Substituting these canonical variables  into the separated relations
\[
2\lambda_i=\left( \hat{u}_i^3\hat{p}_{u_i}^2-a\hat{u}_i-\dfrac{b}{\hat{u}_i^2}-\dfrac{c}{\hat{u}_i^3}\right)=\hat{H}_1\pm\sqrt{\hat{H}_2}\,,\quad i=1,2,
\]
one gets  Hamilton function
\ben
\hat{H}_1&=&\dfrac{u_1^3\Bigl(bu_1u_2+c(3u_1+u_2)\Bigr)p_{u_1}^2}{c(u_1-u_2)}
                +\dfrac{u_2^3\Bigl(bu_1u_2+c(u_1+3u_2)\Bigr)p_{u_2}^2}{c(u_2-u_1)}-a\left(\dfrac{bu_1u_2}{c}+3(u_1+u_2)\right)\nn\\
                \nn\\
&+&\dfrac{b^2(u_1+u_2)}{cu_1u_2}
+\dfrac{b(u_1+2u_2)(2u_1+u_2)}{u_1^2u_2^2}+\dfrac{c(u_1+u_2)(u_1^2+3u_1u_2+u_2^2)}{u_1^3u_2^3}\nn
\en
and the second integral of motion, which is polynomial of sixth order in momenta
\ben
\hat{H}_2&=&
\left(\dfrac{u_1^3u_2^3(u_1^2p_{u_1}-u_2^2p_{u_2})^2}{c(u_1-u_2)^2}-\dfrac{au_1^3u_2^3}{c}
+\dfrac{b^2u_1^2u_2^2}{4c^2}+\dfrac{b(u_1+u_2)u_1u_2}{2c}+\dfrac{(u_1+u_2)^2}{4}
\right)\nn\\
\nn
&\times&
\dfrac{4}{(u_1-u_2)^2}\left(a(u_1-u_2)+b\left(\dfrac{1}{u_1^2}-\dfrac{1}{u_2^2}\right)
+c\left(\dfrac{1}{u_1^3}-\dfrac{1}{u_2^3}\right)-u_1^3p_{u_1}^2+u_2^3p_{u_2}^2\right)^2\,.
\en
Vector field associated with Hamiltonian $\hat{H}_{1}$ is a bi-Hamiltonian vector field with respect to the compatible Poisson brackets
(\ref{poi-11}-\ref{poi-22}).

 If we put   $a=b=0$ and then $c=0$, we obtain a geodesic flow on the plane with an integral of motion of third order in momenta
\bq\label{k-B}\begin{array}{l}
\hat{H}_1=T=\dfrac{ u_1^3(3u_1+u_2)p_{u_1}^2}{u_1-u_2}+\dfrac{u_2^3(u_1+3u_2)p_{u_2}^2}{u_2-u_1}\,,\\
\\
K=\dfrac{u_1^{3/2}u_2^{3/2}(u_1^2p_{u_1}-u_2^2p_{u_2})(u_1^3p_{u_1}^2-u_2^3p_{u_2}^2)}{(u_1-u_2)^2}\,.
\end{array}
\eq
In original Cartesian coordinates these integrals of motion have the form
\[
\hat{H}_1=T=\dfrac{3q_1^2q_2}{2}p_1^2+(q_1^2+6q_2^2)q_1p_1p_2+\dfrac{q_2(5q_1^2+12q_2^2)}{2}\,p_2^2
\]
and
\[
K=q_1^6p_1^3+6q_1^5q_2p_1^2p_2+q_1^4(q_1^2+12q_2^2)p_1p_2^2+2(q_1^2+4q_2^2)q_1^3q_2p_2^3\,.
\]
Construction and classification of all the  integrable geodesic flows on  Riemannian manifolds is  a classical problem in Riemannian geometry \cite{bj04,kiy01,mt01}. We propose to use auto BTs to solution of this problem.  Similar to \cite{kiy01} second integral of motion $K$ (\ref{k-B}) is factorized on two polynomials in momenta but in contrast with \cite{kiy01} factors itself do not commute with $T$. In three dimensional case  similar examples of factorized  integrals of motion are discussed in \cite{ts15}.

\subsection{Case D.}
In this case
\[
f(x)= ax^4+bx^3+cx^2+H_1x+H_2\,,\quad
P(x)=x\left(\dfrac{(x-x_2)y_1}{x_1(x_1-x_2)}+\dfrac{(x-x_1)y_2}{x_2(x_2-x_1)}\right)
\]
and variables $x_{3,4}=\tilde{u}_{1,2}$ are the roots of polynomial
\ben
\dfrac{\psi(x)}{(x-u_1)(x-x_2)}&=&
\left(a-\dfrac{(u_1p_{u_1}-u_2p_{u_2})^2}{(u_1-u_2)^2}\right)x^2
+\left(a(u_1+u_2)+b-\dfrac{u_1^2p_{u_1}^2-u_2^2p_{u_2}^2}{u_1-u_2}\right)\,x\nn\\
\nn\\
&+&\left(
a(u_1^2+u_1u_2+u_2^2)+b(u_1+u_2)+c-\dfrac{u_1^3p_{u_1}^2-u_2^3p_{u_2}^2}{u_1-u_2}\nn
\right)\,.
\en
These coordinates commute with respect to the Poisson brackets
\[
\{u_i,p_{u_j}\}_W=u_i^{-1}\delta_{ij}\,,\qquad  \{u_1,u_2\}_W=\{p_{u_1},p_{u_2}\}_W=0
\]
associated with the Weierstrass change of time.

Using an additional Poisson map
\[
\rho_D:\qquad (u_1,u_2,p_{u_1},p_{u_2})\to (u_1,u_2,u_1^{-1}p_{u_1},u_2^{-1}p_{u_2})\,,\]
we can define canonical variables $\hat{u}_{1,2}$ on $T^*\mathbb R^3$,  which are  the roots of polynomial
\ben
\rho_D\left(\dfrac{\psi(x)}{(x-u_1)(x-x_2)}\right)
&=&
\left(a-\dfrac{(p_{u_1}-p_{u_2})^2}{(u_1-u_2)^2}\right)x^2+
\left(a(u_1+u_2)+b-\dfrac{p_{u_1}^2-p_{u_2}^2}{u_1-u_2}\right)\,x\nn\\
\nn\\
&+&\left(
a(u_1^2+u_1u_2+u_2^2)+b(u_1+u_2)+c-\dfrac{u_1p_{u_1}^2-u_2p_{u_2}^2}{u_1-u_2}\nn
\right)\,,
\en
and the conjugated momenta
\[
\left.\hat{p}_{u_{1,2}}=-\hat{u}_{1,2}^{-1}\,\rho_D(P)\right|_{x=\hat{u}_{1,2}}\,,\quad \rho_D(P)=x\left(\dfrac{(x-u_2)p_{u_1}}{u_1-u_2}+\dfrac{(x-u_1)p_{u_2}}{u_2-u_1}\right)\,,
\]
so that
\[
\{\hat{u}_i,\hat{p}_{u_j}\}=\delta_{ij}\,,\qquad  \{\hat{u}_1,\hat{u}_2\}=\{\hat{p}_{u_1},\hat{p}_{u_2}\}=0\,.
\]
Substituting these variables into the separation relations
\[
2\lambda_i=\Bigl(\hat{u}_i\hat{p}_{u_i}^2-a\hat{u}_i^3-b\hat{u}_i^2-c\hat{u}_i\Bigr)=\hat{H}_1\pm\sqrt{\hat{H}_2}\,,\quad i=1,2,
\]
one gets  Hamilton function
\ben
\hat{H}_1&=&\dfrac{u_1(2u_1+u_2)p_{u_1}^2}{u_1-u_2}+\dfrac{u_2(u_1+2u_2)p_{u_2}^2}{u_2-u_1}
-a(u_1+u_2)(2u_1^2+u_1u_2+2u_2^2)\nn\\
\nn\\
&-&b(2u_1^2+3u_1u_2+2u_2^2)-2c(u_1+u_2)
\en
and the second integral of motion, which is polynomial of fourth order in momenta
\[
\begin{array}{rcl}
\hat{H}_2&=&u_1^2u_2^2\Bigl(\frac{(p_{u_1}-p_{u_2})^3\bigl((3u_1+u_2)p_{u_1}+(u_1+3u_2)p_{u_2}\bigr)}{(u_1-u_2)^3}
-\frac{4c(p_{u_1}-p_{u_2})^2}{(u_1-u_2)^2}\Bigr.\nn\\
\nn\\
&-&\frac{2a\bigl(3(u_1^2+u_2^2)(p_{u_1}^2+p_{u_2}^2)-4(u_1^2+u_1u_2+u_2^2)p_{u_1}p_{u_2}\bigr)}{(u_1-u_2)^2}
-\frac{2b(p_{u_1}-p_{u_2})\bigl((u_1+3u_2)p_{u_1}-(3u_1+u_2)p_{u_2}\bigr)}{(u_1-u_2)^2}\nn\\
\nn\\
&+&\Bigl.a^2(3u_1^2+2u_1u_2+3u_2^2)+a(2b(u_1+u_2)+4c)-b^2\Bigr)\,.
\end{array}
\]
Vector field associated with Hamiltonian $\hat{H}_{1}$ is a bi-Hamiltonian vector field with respect to the compatible Poisson brackets
(\ref{poi-11}) and (\ref{poi-22}).

If we put $a=b=0 $ and then $c=0$,  we obtain a geodesic flow with an integral of motion, which is polynomial of fourth order in momenta
\ben
\hat{H}_1=T&=&\dfrac{u_1(2u_1+u_2)p_{u_1}^2}{u_1-u_2}+\dfrac{u_2(u_1+2u_2)p_{u_2}^2}{u_2-u_1}\,,\nn\\
K&=&\dfrac{u_1^2u_2^2(p_{u_1}-p_{u_2})^3\Bigl((3u_1+u_2)p_{u_1}+(u_1+3u_2)p_{u_2}\Bigr)}{(u_1-u_2)^3}\,.
\nn
\en
This integrable system is superintegrable system with one more integral of motion, which is  a polynomial of  third order in momenta
\[
J=\dfrac{u_1u_2(p_{u_1}-p_{u_2})^2(u_1^2p_{u_1}-u_2^2p_{u_2})}{(u_1-u_2)^3}\,,
\]
so that
\[
\{T,K\}=0\,,\qquad \{T,J\}=0\,,\qquad \{J,K\}\neq 0\,.
\]
In original Cartesian coordinates these integrals of motion are
\[
T=q_2\left(\dfrac{p_1^2}{2}+p_2^2\right)+\dfrac{q_1p_1p_2}{2}\,,\quad
K=\dfrac{(q_1^2-q_2^2)p_1^4}{4}+\dfrac{q_1q_2p_1^3p_2}{2}\,,\quad
J=(q_1p_1+2q_2p_2)p_1^2\,.
\]
Using additional canonical transformation we can reduce polynomial $T$ to the following form \[T=m_1(q_1,q_2)p_1^2+m_2(q_1,q_2)p_2^2\]
and obtain new example of superintegrable system with the position dependent masses, see  \cite{ran16} and references therein.

\section{Conclusion} We prove that auto B\"{a}cklund transformation of Hamilton-Jacobi equation which represents symmetry of the level manifolds can be also useful  in classical mechanics as the auto B\"{a}cklund transformation associated with addition law in Jacobian.

 In particular,  we prove that geodesic flow with diagonal in parabolic coordinates metric
\bq\label{gen-metr}
\hat{\mathrm g}^{(km)}=\left(
                 \begin{array}{cc}
                   \dfrac{(ku_1+u_2)u_1^m}{u_1-u_2}& 0 \\
                   0 & \dfrac{(ku_2+u_1)u_2^m}{u_2-u_1} \\
                 \end{array}
               \right)
\eq
is integrable for $k=3;\,m=3,4$ and for $k=2;\,m=1$. The corresponding integrals of motion
\[
K=(u_1^2p_{u_1}-u_2^2p_{u_2})\cdot\dfrac{u_1^{m/2}u_2^{m/2}(u_1^mp_{u_1}^2-u_2^mp_{u_2}^2)}{(u_1-u_2)^2}\,,\qquad m=3,4
\]
and
\[
K=(u_1^2p_{u_1}-u_2^2p_{u_2})\cdot\dfrac{u_1^{m}u_2^{m}(p_{u_1}-p_{u_2})^2}{(u_1-u_2)^3}\,,\qquad m=1
\]
are polynomials of the third order in momenta  with a common factor. It allows us to suppose, that there are similar integrable systems for other values of $k$ and $m$ in (\ref{gen-metr}), see discussion in \cite{bj04,kiy01,mt01}. Indeed, substituting
\[
T=\sum_{i,j=1}^2 \hat{\mathrm g}_{ij}^{(km)}(u)p_{u_i}p_{u_j}
\]
and
\[
K=(u_1^2p_{u_1}-u_2^2p_{u_2})\cdot\left(f(u_1,u_2)p_{u_1}^2+g(u_1,u_2)p_{u_1}p_{u_2}+h(u_1,u_2)p_{u_2}^2\right)
\]
into the equation $\{T,K\}=0$ and solving the resulting system of PDE's with respect to functions $f,g$ and $h$ we can prove the following proposition.
\begin{prop}
Metric $\hat{\mathrm g}^{(km)}$ (\ref{gen-metr}) yields integrable geodesic flow on the plane at $m=1,k=2$ and
\[
m=3\,,\qquad k=\pm1,3,\dfrac12\,,\qquad\mbox{and}\qquad m=4\,,\qquad k=\pm1,\pm3,-\dfrac35,-\dfrac{1}{7},\dfrac{1}{5},\dfrac{1}{2}\,.
\]
\end{prop}
In the similar manner we can study common properties of the obtained variables of separation, compatible Poisson brackets and recursion operators. It can be useful for investigation of other integrable systems with integrals of motion of third, fourth and sixth order in momenta. In particular we hope to use obtained experience for the study of Toda lattice associated with $G_2$ root system.

 Starting with other well-known separable  Hamilton-Jacobi equations on the plane, sphere, ellipsoid,
 Lorentzian  space, de Sitter and anti-de Sitter spaces we can also obtain  new integrable systems with polynomial integrals of motion. We plan to describe some of these systems in the forthcoming publications.

The work was supported by the Russian Science Foundation (project  15-11-30007).

\end{document}